# Sliding-ferroelectric control of dipolar and quadrupolar excitons in R-stacked $WSe_2$

Asmund Ottesen[1,2*], Shun Feng[1,2,3*§], Edoardo Lopriore[1,2], Yaroslav Zhumagulov[4], Kenji Watanabe[5], Takashi Taniguchi[6], Oleg V. Yazyev[4], Andras Kis[1,2§]

[1]*Institute of Electrical and Microengineering, École Polytechnique Fédérale de Lausanne (EPFL), CH-1015 Lausanne, Switzerland*
[2]*Institute of Materials Science and Engineering, École Polytechnique Fédérale de Lausanne (EPFL), CH-1015 Lausanne, Switzerland*
[3]*Institute of Applied Physics and Materials Engineering, University of Macau, Macao SAR, China*
[4]*Institute of Physics, Ecole Polytechnique Fédérale de Lausanne (EPFL), CH-1015 Lausanne, Switzerland*
[5]*Research Center for Electronic and Optical Materials, National Institute for Materials Science, 1-1 Namiki, Tsukuba 305-0044, Japan*
[6]*Research Center for Materials Nanoarchitectonics, National Institute for Materials Science, 1-1 Namiki, Tsukuba 305-0044, Japan*

[*]*These authors contributed equally.*
[§]*Correspondence should be addressed to: Andras Kis (andras.kis@epfl.ch) and Shun Feng (shunfeng@um.edu.mo)*

## ABSTRACT

A fundamental property of interlayer excitons is their dipolar character, enabling electrical tunability, enhanced exciton-exciton interactions and extended lifetime. It is important for studying collective phenomena and realizing novel optoelectronic devices. However, in homo- or heterostructures based on two-dimensional semiconductors, the magnitude and orientation of the dipole moment are commonly locked due to spatially confined charges and nature of the band transition. Here, we use gate-dependent photoluminescence of few-layer R-$WSe_2$ to harness sliding ferroelectricity for controlling the exciton dipole configuration of emerging layer-hybridised excitons. The sliding-ferroelectric control enables a pronounced on/off switching of the diffusive transport of hybrid dipolar excitons, while further facilitating gradual and abrupt dipole reconfigurations with emerging quadrupolar excitons. Our optical probe reveals intermediate states of vertically aligned interfacial polarisation with magnitudes that confirm cumulative polarisation in multilayers. Overall, our work presents sliding ferroelectricity as a powerful tool to tune interlayer coupling and layer hybridisation. These results represent a step toward ferroelectric-controlled exciton interconnects and novel multi-bit optical memories.

## INTRODUCTION

Long-lived exciton states represent attractive candidates for next-generation optoelectronics[1–5]. In particular, the out-of-plane electric dipole characteristic of interlayer excitons enables a combination of electrical and optical control routes, holding strong potential for non-linear optics[6,7] and quantum simulation[8]. The electrical control of these quasiparticles is achieved by applying a vertical electric field to tune their energy and interactions via the quantum-confined Stark effect[2,3,5]. In particular, tunable electron-hole wavefunction overlap directly affects radiative recombination and exciton lifetime[5,9]. Their dipolar nature has a relevant role within the interplay of their interactions[6,10,11], which in turn dictates their propagation and has enabled switching of excitonic transport via field-induced potential landscapes[3,4,12]. Thus, controlling the effective dipole moment of interlayer excitons is a key objective for the development of excitonic devices and for tuning exciton interactions in collective phenomena, such as excitonic insulator states[8,13–18].

Van der Waals heterostructures based on transition metal dichalcogenides (TMDs) are a versatile platform for hosting interlayer excitons, where two or more vertically stacked layers enable strong interlayer coupling. By varying the material composition, atomic registry or twist angle, new materials can be designed with tailored excitonic properties that arise from interlayer coupling, such as hybrid[9,19] and moiré excitons[20,21].

Quadrupolar excitons (QX) have recently emerged as a novel type of layer-hybridised excitons with field-tunable wavefunctions and dipole moment[22–26]. The QX can be regarded as a superposition of two oppositely oriented dipolar excitons hybridised via interlayer carrier tunnelling, yielding a distinct non-linear quadratic Stark shift. Upon an applied field, the effective dipole moment gradually evolves from vanishing (symmetric quadrupolar wavefunction) to that of a dipolar exciton (layer-polarised wavefunction). This enables gradual tuning of exciton-exciton interactions between quadrupolar and dipolar regimes. In contrast, additional abrupt, hysteretic control of the excitonic dipole could enable switching between programmable states for low-power optoelectronics.

Rhombohedral ($R$) stacked TMDs have recently emerged as a promising material platform for achieving fast ferroelectric switching at a low energy cost for future non-volatile memory applications[27–31]. The atomic configuration induces a spontaneous electric polarisation that can be reversed with field-controlled layer sliding, giving rise to sliding ferroelectricity[32–35]. This opens the door to using the intrinsic polarisation as a new degree of freedom, combined with the already rich electrical and optical properties of TMD structures[36]. Ferroelectric hysteresis is then encoded into the excitonic properties for exciton-based optical memories and reconfigurable optoelectronic devices[37]. With fast switching dynamics of sliding ferroelectricity and electrical field-effect control of excitons coupled within the same device, it's possible to engineer optoelectronics based on 2D semiconductors beyond current state of the art[38,39]. Previous efforts have mainly focused on probing intralayer excitons; however, their lack of a dipole moment means the exciton energy is unaffected by an applied field. In contrast, interlayer excitons, with their field-tunable energy and spatial extent, offer an alternative way to probe changes in atomic registry and the local dielectric environment[27,29,32]. Furthermore, they provide direct access to the polarisation strength via their Stark shift.

Significant attention has been directed towards natural 3R-$MoS_2$[27–31,40], however artificial stacking with marginal twist also offers a route for sliding ferroelectricity and sizable domains[32–34,41,42]. Progress has centred around bilayers with two-bit ($\pm P$) polarisation switching, while scaling beyond bilayers provides an opportunity for multi-ferroelectricity - an unparalleled feature of layered ferroelectrics. For a homostructure of $n$ layers comprised of rhombohedral interfaces, the number of distinct cumulative configurations is then given by $2^{n-1}$, enabling the possibility for multi-bit information storage[31,34,43,44]. However, the interlayer excitons in rhombohedral homostructures are less explored. In particular, the impact of interfacial polarization switching on the interlayer exciton hybridization and dipolar character in multilayer homostructures still remains elusive.

Here, we optically probe the Stark shift of emerging hybrid-dipolar and quadrupolar excitons in few-layer R-$WSe_2$ and demonstrate unprecedented sliding-ferroelectric control over the exciton dipole. In bilayer R-$WSe_2$, we observe large (>10 $\mu m^2$) switchable domains of $\pm P$ polarisation. This allows us to abruptly modify the orientation of hybrid dipolar excitons and consequently, their diffusive transport. In trilayer and tetralayer R-$WSe_2$, we observe multiple signatures of interfacial sliding and intermediate polarisation states. With quadrupolar excitons, we gradually tune the effective dipole moment by applying an external field and achieve abrupt switching via interfacial sliding. From these results, we experimentally determine the polarisation strength and confirm cumulative polarisation with layer scaling by analysing the field offset in the exciton Stark shift. The study of layer-hybridised excitons with controllable dipole tuning and hysteretic ferroelectricity sets the stage for future multi-bit optical memories based on van der Waals materials.

## RESULTS

### Excitonic sensor for sliding ferroelectricity in R-stacked $WSe_2$

We fabricate artificially R-stacked (marginal 0°-twist) few-layer $WSe_2$ with terraces of 2, 3 and 4L, encapsulated with hBN dielectrics in a dual-gated architecture (Figure 1a). We provide more details on the device in Supplementary Note 1 and the the Methods section. This geometry allows us to selectively apply either electric field $E_z$ or electrostatic doping $n_e$ (Supplementary Note 2). As we contact two layers with a rhombohedral interface, atomic reconfiguration leads to the formation of spatial domains with out-of-plane electrical polarization facing either up (+$P$) or down (-$P$), as illustrated in Figure 1b. We confirm the formation of polarized domains in a representative R-stacked $WSe_2$ bilayer using Kelvin Probe Force Microscopy (KPFM, Supplementary Note 3). Further, by vertically aligning more rhombohedral interfaces via layer scaling, the polarisation from each interface accumulates, leading to a net polarisation with unique interfacial combinations[34,44], as shown in Figure 1c. Since the aligned polarisations are largely decoupled, the lateral sliding at each interface can be induced at different field magnitudes, enabling electrical control over intermediate polarisations and multi-ferroelectricity[27,31,43].

We show hysteresis loops for 2, 3 and 4L R-$WSe_2$ in Figure 1d, representing the switching pathways in our device. Here, sliding ferroelectricity leads to a hysteresis with a 'memory window' when sweeping the electric field between opposite directions. The interfacial sliding is apparent from the discrete steps in net polarisation of $|2P|$ at a given coercive field. The size of the 'memory window' depends on the difference in coercive fields, while multilayers form several memory windows, useful for multi-bit programmable device operations[39].

Interlayer excitons, which spatially extend across several layers, present an excellent optical and non-destructive probe for studying cumulative ferroelectricity and switching pathways in few-layer R-TMDs. We perform photoluminescence spectroscopy (PL) to probe phonon-assisted momentum-indirect interlayer transitions in our terraced few-layer R-$WSe_2$, as seen in Figure 1e. Here, we observe a general emission energy redshift for increasing layers due to band gap closure[45,46]. Furthermore, the emission intensity decreases with layer scaling because thicker samples have reduced radiative recombination.

Beyond bilayers, we find that the lowest energy transition changes from KΛ to ΓΛ in momentum space, as confirmed by our DFT calculations (Supplementary Note 4). We further extract the wavefunction distributions in real space, to understand the layer (de) localisation of charges, as shown in Fig. 1f-g. For bilayer R-$WSe_2$, the spatial distribution indicates localised holes in the K-valley and delocalised electrons in the Λ-valley, resulting in hybrid dipolar excitons (hIX), consistent with previous reports[47–50]. In our calculations, we consider the sliding-induced atomic configuration with $+P$, which introduces a slight asymmetry in the electron wavefunction distribution. On the other hand, for trilayer R-$WSe_2$ ($0P$), we predict fully delocalized electron and hole wavefunctions across all layers, stemming from the more energetically favourable ΓΛ transition. This symmetric delocalisation at the Γ(Λ)-valley[51,52] leads to the formation of emerging quadrupolar excitons (QXs) formed via interlayer charge tunneling[22,24,25,53]. In the following sections, we use hIX and QX to sense interfacial sliding and polarisation switching.

**Optical probing of sliding ferroelectricity in bilayer R-$WSe_2$**

Sliding ferroelectricity in our R-stacked bilayer $WSe_2$ manifests as a hysteresis in the excitonic features when sweeping an applied electric field $E_z$. We probe the low-energy exciton $hIX_{K\Lambda}$ with photoluminescence spectroscopy (PL) while sweeping the field from negative to positive (forward) and positive to negative (backward). We first consider the forward sweep in Figure 2a. Here, we observe a linear Stark shift centered around zero electric field. We extract a dipole moment ($\Delta E = -E_z d$) of $0.3 \pm 0.014$ nm (Supplementary Note 5), which indicates a hybrid dipolar exciton with a momentum-indirect transition featuring delocalization of one or more charges. Similar dipole sizes were observed in natural and twisted $WSe_2$ homobilayers and R-$MoSe_2$ with marginal twist[27,32,48]. We attribute the intensity variations close to zero field to unintended doping, which can drastically change the intensity (Supplementary Note 6).

We also observe a crossover at $-0.12\ \mathrm{V/nm}$ between two sets of oppositely shifting peaks that exchange oscillator strength. Each exhibits a permanent dipole moment of equal magnitude, however their dipole is oppositely oriented. We attribute this feature to a crossover of the layer-dependent valence band maxima (VBM), as seen in our DFT-calculated band structure in Supplementary Note 4. While this is a characteristic feature for TMD homobilayers[27,32,48], such a crossover in R-stacked samples is effectively offset from zero field due to intrinsic polarisation. We note that the oppositely oriented dipoles exhibit negligible coupling when tuned in resonance.

Notably, we observe an abrupt energy shift at $E_{c,f} = 0.068\ \mathrm{V/nm}$ followed by a Stark shift indicating a flip in the dipole orientation. This stepwise modulation points to sliding ferroelectricity, in which an out-of-plane field drives a lateral displacement that flips the internal polarisation. Comparing the energy shift with the calculated band diagram, we confirm a polarisation reversal from -*P* to +*P* (Supplementary Note 7). We observe a second VBM crossover at $0.12\ \mathrm{V/nm}$, matching the crossover initially observed at $-0.12\ \mathrm{V/nm}$, confirming the reversal of the internal polarisation.

During the backward sweep (Figure 2b), the excitonic energy dispersion initially mirrors that of the forward sweep, since the polarisation remains in the +*P* state. However, as we sweep the field to $E_z < E_{c,f}$, the dispersion continues in the +P state. Eventually, we observe another abrupt energy shift and dipole orientation flip at $E_{c,b} = -0.045\ \mathrm{V/nm}$, switching back to -*P* polarisation. The complete field-sweep loop thus displays both abrupt spectroscopic features and reversible hysteresis, corroborating our assignment of sliding ferroelectricity.

To further strengthen support for this interpretation, we measure the field-dependent reflectance spectra of intralayer excitons (*X*). In Figure 2c-d we show the first derivative of the reflectance contrast (RC) to highlight the spectral features. We observe two main distinct peaks, which we identify as intralayer excitons originating from top/bottom layers ($X_t, X_b$). These momentum-direct intralayer excitons are non-degenerate due to the bandgap modification of the two metastable atomic configurations in R-stacked TMDs[27]. The 14.9 meV energy separation between the peaks is consistent with other reports[49,50]. We further observe anti-crossings between intra- and interlayer excitons from spin-conserved hole tunnelling (Supplementary Note 8). The polarisation switching here emerges as an intensity re-distribution between the two observed peaks $(X_b, X_t)$ at the same coercive fields $(E_{c,f}, E_{c,b})$ as found in PL. We further report abrupt features in the PL emission of intralayer excitons (Supplementary Note 9). We confirm the KΛ transition by measuring the doping dependent reflectance (Supplementary Note 10).

To emphasise the hysteresis in both field-dependent PL and RC, we plot exemplary spectra at fixed electric fields in Figure 2e-f. These highlight the energy shift to $hIX_{K\Lambda}$ in PL spectra and intensity redistribution between $X_b, X_t$ in RC spectra, showing distinguishable spectral features between $\pm P$ polarisation states.

We quantify the hysteretic energy dispersion by extracting the $hIX_{K\Lambda}$ energy shift in both forward and backward sweeps, as seen in Figure 2g. The difference between the coercive fields gives the memory window ($\Delta E_c$), yielding a size of $0.112\ \mathrm{V/nm}$. The reversible nature is highlighted by the overlapping Stark shift for fields $E_z > E_{c,f}$ and $E_z < E_{c,b}$. From the field difference between the observed VBM crossovers in forward and backward sweeps, we find a total field offset of $0.23\ \mathrm{mV/nm}$ corresponding to $2P$. Given an interlayer distance of 0.6 nm we estimate the interlayer potential $P = 69$ mV. Furthermore, from linear extrapolation of $hIX_{\mathrm{K}\Lambda}$ we can extract the valence band splitting, yielding $\Delta_{VBM}= 66.3\ \pm 3.4\ \mathrm{meV}$ (Supplementary Note 11).

We identify four distinct combinations characterised by the exciton dipole orientation and its (anti-)alignment with the applied field, schematically represented with insets in Figure 2g. This provides an additional degree of freedom to engineer dipolar exciton diffusion with a hysteretic approach, discussed in the following section.

**Ferroelectric switching of exciton diffusion**

With our artificially R-stacked $WSe_2$ bilayer, we realise sample-wide polarisation domains with electric-field hysteresis, as seen in Figure 3a. We use spatial PL mapping to locate the domain wall boundaries and trace their propagation using the $hIX_{\mathrm{K\Lambda}}$ peak energy as a probe, in accordance with previous reports on 3R-$MoS_2$[28,29,31,40]. With various training fields ($E_{tr}$), we precisely track intermediate domain wall positions as the domains expand/shrink across the bilayer. While mapping, we apply a small measuring field ($E_{meas} = -0.04\ \mathrm{V/nm}$) which separates the peak energy of $hIX_{\mathrm{K\Lambda}}^{\downarrow}$ and $hIX_{\mathrm{K\Lambda}}^{\uparrow}$ by roughly $20$ meV (Fig. 2g) to clearly distinguish the $\pm P$ polarization. As a result, we observe large and uniform (tens of $\mu m^2$) polarization domains and field-tunable domain wall motion. We flip the polarization across most of the bilayer flake (Supplementary Note 12). However, some smaller regions remain locked into a constant polarization, likely caused by mesoscopic strain anchoring at flake edges[29,31]. We attribute the highly uniform energy distributions within each domain to the ultraclean, bubble-free sample quality (Supplementary Note 1). We confirm reversible domain wall motion at room temperature across repeated cycles of PL maps, suggesting that pinning potentials exceed the thermal energy[29]. This lays a foundation for spatially programmable interlayer potentials, using training fields to control dipolar exciton diffusion.

Based on the PL mapping, we choose a position where the domain wall moves across the entire probing spot without intermediate pinning to reliably measure discrete switching of exciton diffusion. We repeat the field-dependent PL of $hIX_{\mathrm{K\Lambda}}$ to determine the coercive fields in this spot, as seen in Figure 3b,c. Here, we find low coercive fields of $E_{c,b} = -0.058\ \mathrm{V/nm}$ and $E_{c,f} = 0.055\ \mathrm{V/nm}$ and a qualitatively similar Stark shift as in Figure 2a-b. Here, the hIX exciton initially redshifts along the sweep direction and then blueshifts after the coercive field, for both sweep directions. We therefore conclude that the dipole moment initially aligns with the applied field and then flips to become anti-aligned. We simultaneously measure spatial PL images (Supplementary Note 13) to quantify the effective diffusion area of $hIX_{\mathrm{K\Lambda}}$, which we show in Figure 3d,e. Exactly at the coercive fields, the diffusion area sharply decreases with a stepwise modulation, effectively switching from an 'on' to an 'off' diffusive state. We explain this by considering the interlayer exciton lifetime with field-tunable electron-hole wavefunction overlap[5,6]. As the exciton dipole switches from being aligned to anti-aligned with respect to the applied field, this sharply increases the overlap of electron and hole wavefunctions, shortening the radiative lifetime and thereby the effective diffusion length. Our measurements of exciton lifetime as a function of field, including polarisation switching, support this picture (Supplementary Note 14). Figure 3f, g also supports this by showing PL intensity across the field sweeps. In particular, we observe sharp increases in PL intensity at the coercive fields, as the probability of radiative recombination is promoted[5]. We find diffusion modulation switching of 8-15%, which can potentially be enhanced by increasing the coercive fields, as the lifetime is expected to be linearly proportional to the field.

**Multi-polarization switching in 3L- and 4L-$WSe_2$**

Lastly, we demonstrate sliding-ferroelectric control of emerging quadrupolar excitons (QXs) in trilayer (3L) and tetralayer (4L) R-$WSe_2$ (Figure 4a,e). Here, we experimentally observe QXs via a momentum-indirect ΓΛ transition (Figure S4). Our QX uniquely features symmetric delocalisation of both electron and hole (Figure 1g), which defines an electric quadrupole and yields a quadratic Stark shift as a function of vertical electric field[23,24,53]. Such quadratic field polarizability can be phenomenologically regarded as a pair of anti-parallel dipoles coupled with a finite strength $t_0$. We further fabricate natural 2H-$WSe_2$ multilayers (3, 4 and 5L) and similarly confirm a quadratic Stark shift from the same ΓΛ-transition (Supplementary Note 15). Here, the absence of internal polarisation leads to a field-neutrality point at zero field.

However, with rhombohedral stacking and internal polarisation, a finite field is required to realise the symmetric quadrupolar wavefunction and consequently a vanishing dipole moment. Importantly, the large spread of the electron and hole wavefunctions makes the QX sensitive to atomic reconfiguration and polarisation switching at each interface. Thus, QXs serve as a unique probe for sensing sliding multiferroelectricity via a predictable quadratic Stark shift, while also offering a novel opportunity for complex dipole engineering.

We demonstrate the field-tunability of QX in trilayer R-$WSe_2$ (Figure 4a) comprising two ferroelectric interfaces, thus yielding three possible net polarisations of -2$P$, 0$P$, +2$P$. In the forward sweep (Figure 4a, left), we resolve a characteristic quadratic energy shift from -0.17 to ~0.07 V/nm, followed by two discontinuities in the energy dispersion at coercive fields of 0.072 V/nm and 0.08 V/nm, respectively. We thus assign the three polarisation regimes initially at $-2P$, then intermediate $0P$, and lastly $+2P$. In contrast, in the backward sweep (Figure 4a, right), the net polarisation initialises at +2$P$, and we now resolve the polarisations +2$P$, 0$P$, -2$P$, with the coercive fields at -0.047 V/nm and -0.057 V/nm. We extract the field-dependent energy shift of the 3L-QX using a representative phonon replica peak (Supplementary Note 16) and bunch the extracted energies into their distinct net polarisation (Figure 4b). We then separately fit the bunched energy shifts with the hyperbolic function $\Delta E(E_z) = -\sqrt{(t_0)^2 + (eE_z \cdot d_0)^2}$, where $d_0$ is the dipolar constituent and $t_0$ is the coupling strength of the anti-aligned dipole pair. At high effective field, we can estimate the starting point of $d_0$ from the linear dispersion regime (Supplementary Note 17). We obtain values of $d_0 = 0.28 \pm 0.01$ nm and $t_0 = 37.5 \pm 2.8$ meV, averaged between polarisations $-2P$ and $+2P$. The hyperbolic model quantitatively captures the experimentally observed energy shifts. We further calculate the effective dipole moment as a function of external field (Figure 4c). Interestingly, we resolve two types of evolution of the dipole moment as a function of field. The first is a linear trend with constant net polarisation. This can be understood as a field-induced layer polarisation of the QX where the dipole moment is expected to approach an asymptote value close to $d_0$. We do not resolve this dipole saturation, which may allow further tuning. The second is the abrupt change of dipole moment occurring at each coercive field. This provides an anomalous tuning knob for engineering the dipole moment from positive to near zero, then to negative, all in a narrow window near zero external field. We summarise the representative regimes in Figure 4d (i, ii and iii), denoting the dipole moment and corresponding ferroelectric polarisation.

Further, we investigate the multi-ferroelectricity in tetralayer R-$WSe_2$, with possible net polarisation of $-3P$, $-1P$, $+1P$, $+3P$. We conduct forward and backward field sweeps (Figure 4e) and trace the field-dependent evolution of 4L-QX. With the forward sweep as an example, the quadratic Stark shift now displays a larger curvature, yielding $d_0 = 0.53 \pm 0.08$ nm and $t_0 = 64.8 \pm 5.9$ meV in the $-3P$ polarization. The enhancement of $d_0$ and $t_0$ is explained by an expected larger wavefunction spread across four layers, indicating an effective layer-engineering strategy for QX (Supplementary Note 18). We resolve three abrupt jumps in the forward and backward sweeps, respectively. Analysing the experimental energy dispersion, we can assign three of the four possible net polarisations, namely $-3P$, $-1P$, $+3P$ based on the field offset of effective field neutrality points (Figure 4f). Note that for the switching from $-1P$ to $+1P$, despite an intensity jump in the density plot, we do not resolve the offset in the energy dispersion. This may be due to complex domain migration across multiple sliding interfaces, where opposite migrations compensate each other. Our far-field measurements therefore probe collective multiple-domain migration. At higher fields, the interfaces become fully polarised and aligned with the external field direction, yielding clear offset features for $+3P$ ($-3P$). We proceed to extract the dipole moments for 4L R-$WSe_2$ and similarly achieve both gradual and abrupt tuning of the dipole moment due to field polarisation and ferroelectric switching, respectively (Figure 4g). We schematically summarize the rich dipole engineering of tetralayer R-$WSe_2$ across different regimes in Figure 4h. Remarkably, we achieve gradual tuning of the dipole moment from 0.6 to -0.2 nm, as well as abrupt switching of the dipole magnitude and orientation from i to ii and from ii to iii. Note that for both 3L and 4L devices, ferroelectric switching materialises near zero external gating, which is key to limiting power consumption. For the trilayer, the intermediate state can be stabilised with zero net polarisation at zero field. This could switch exciton-exciton interactions from a symmetric quadrupolar ($0P$) to a layer-polarised dipolar wavefunction ($\pm 2P$) without sustaining an electric field, showing potential for low-power excitonic and optoelectronic devices.

From the hyperbolic fits, we extract accurate field offsets (Figure 4b, f) for determining the experimental cumulative potential of trilayer ($2P$) and tetralayers ($3P$). We summarise the layer-dependent scaling of polarisation potentials optically probed with hybrid dipolar and quadrupolar excitons in Supplementary Note 19. We also plot hysteresis loops for 2, 3, and 4L in Figure S24 in Supplementary Note 20, using the experimentally extracted polarisation potential and coercive fields. This quantitatively demonstrates layer scaling of multi-ferroelectric switching and cumulative polarisation using hybrid interlayer excitons as optical probes in layered R-$WSe_2$.

## CONCLUSIONS

Overall, we demonstrate sliding multi-ferroelectricity as a powerful tool to engineer the dipolar character of hybrid interlayer excitons in R-TMDs. These findings open avenues to optically probe multiple-domain migration and cumulative polarisation scaling in stacked heterostructures. Furthermore, these prospects could enable exploratory optoelectronic device concepts, such as an excitonic analogue of a sliding-ferroelectric field effect device[36,39,54], towards excitonic memories and synapse[55] prototypes.

## ACKNOWLEDGEMENTS

We acknowledge the help of Z. Benes (EPFL Center of MicroNanoTechnology (CMI)) with electron-beam lithography. This project has received funding from the European Union's Horizon 2020 research and innovation programme under grant agreements No. 956813 (Marie Curie Sklodowska ITN network "2-Exciting") and SNSF grant No. 215089. K.W. and T.T. acknowledge support from JSPS KAKENHI (Grant Numbers 19H05790, 20H00354 and 21H05233). Y.Z. and O. V. Y. acknowledge support by the Swiss National Science Foundation (grant No. 224624). Ab initio calculations have been performed at the Swiss National Supercomputing Centre (CSCS) under Project No. lp96 and the facilities of Scientific IT and Application Support Center of EPFL.

## AUTHOR CONTRIBUTIONS

A.K. initiated and supervised the project. A.O. fabricated the R-$WSe_2$ devices. A.O. and S.F. performed the measurements with assistance from E.L. K.W. and T.T. grew the h-BN crystals. Y.Z. and O. V. Y. developed the theoretical framework and performed calculations. A.O., S.F., and A.K. wrote the manuscript with contributions from all authors.

## COMPETING INTERESTS

The authors declare no competing financial interests.

## METHODS

### Device fabrication

We fabricate our devices by encapsulating few-layer rhombohedral tungsten diselenide ($WSe_2$) with hexagonal boron nitride (hBN) gate dielectrics. We exfoliate the van der Waals materials hBN and $WSe_2$ on $SiO_2$ and PDMS (Gel-Pak) substrates, respectively. We identify flake thicknesses by colour contrast under an optical microscope. To transfer the layers, we first attach a thin polycarbonate (PC) layer to a PDMS stamp for sequential pick-up transfers. We use the 'tear-and-stack' transfer method, which stacks layers with marginal twist[56,57,32,41]. Here, our twist-angle resolution is ~ 0.1 degree. We release the stack onto a bottom hBN and graphite gate, prefabricated by a separate PC transfer, using progressive adhesion at temperatures above 150° to melt the PC layer. We then detach the PDMS and cool the stack to room temperature. To remove PC, we use chloroform and high vacuum ($10^{-6}$ mbar) annealing at 340 °C. We then deposit a 4 nm platinum (Pt) top gate by electron-beam lithography and evaporation. Lastly, in a second e-beam lithography and evaporation step, we deposit 2 nm/80 nm titanium/gold (Ti/Au) to contact the top gate, bottom gate and contact electrodes.

**Optical measurements and electrostatic gating**

We performed all the optical measurements under vacuum in a He-flow cryostat at 5K unless otherwise explicitly stated. We use a custom-built confocal microscope setup for both photoluminescence and reflectance spectroscopy. The objective (Olympus LCPLN50XIR) is used for both excitation and collection and has a working distance of 4.5mm and a numerical aperture of 0.65. For PL measurements we excite with a continuous wave 640 nm laser (PicoQuant, LDH-IB-640-M) that has a diffraction limited focus spot of ~1µm. The emitted light is directed to a spectrometer (Princeton Instruments SpectraPro 500) with a grating of 300 g/mm and a CCD camera (Princeton Instruments, Blaze 400-HR/HRX). We use a long pass (LP) of 700 nm to filter out the laser excitation. For reflectance spectroscopy we replace the laser excitation with a Tungsten-Halogen white-light excitation (Thorlabs SLS201L). To acquire PL spectra and spatial PL images simultaneously, we insert a 50/50 beamsplitter in the collection path which directs the emitted light to both the spectrometer and a CCD camera (Andor Ixon). To selectively image the diffusion of 2L-hIX, we use both a LP 775nm filter and a short pass 900nm filter before the CCD camera. This effectively removes the laser excitation and $WSe_2$ intralayer exciton emission. For time resolved PL measurement, the same 640 nm laser diode is used in pulsed mode to produce pulses of 160ps at a repetition rate of 80 MHz. We cut emitted photons from the sample with a LP 775nm filter and fiber couple into a single photon avalanche diode (SPAD, MPD PDM). The time delay of each photon is then tracked with a time-correlated photon-counting module (TCPC, PicoQuant, PicoHarp 300) having a resolution of 12 ps r.m.s. The setup is time-limited by the resolution of the SPAD which is 50 ps.

For gating, we apply voltages on both gates using the same source meter (Keithley 2636B) while the sample grounded. For PL diffusion with electric field control, we first apply a high training field to drive the polarization into either $\pm P$. We then sweep the field in the opposite direction while simultaneously acquiring PL spectra and spatial images.

The reflectance contrast is calculated with the following relation $RC = (R - R_{bg})/R_{bg}$, where $R$ and $R_{bg}$ are reflectance spectra recorded with and without the R-$WSe_2$, respectively.

**Ab-initio calculations**

As a preliminary step in our theoretical analysis, we performed DFT calculations on R-$WSe_2$ using the GPAW code[58] in plane-wave mode. We employ a kinetic-energy cutoff of 500 eV and the local-density approximation (LDA) for exchange correlation. Brillouin-zone sampling used a Γ-centered 12 × 12 × 1 mesh. The periodic replicas of R-$WSe_2$ were separated by 10 Å of vacuum along the $z$ direction. Structural optimization was performed with the LBFGS algorithm, relaxing the in-plane lattice parameters and ionic positions while keeping the out-of-plane lattice parameter and all cell angles fixed, until the maximum residual force was below $10^{-3}$ eV Å$^{-1}$. To understand the optical response of the R-$WSe_2$, we evaluated the exciton Hamiltonian within the Bethe–Salpeter equation (BSE)[59]. In the electron–hole basis:

$$(E_c - E_v)A_{vc}^S + \sum_{v'c'} K_{vc,v'c'}\, A_{v'c'}^S = \Omega_S A_{vc}^S \tag{1}$$

where $E_c$ and $E_v$ are the single-particle conduction-band and valence-band energies, respectively; $K_{vc,v'c'}$ denotes the electron–hole interaction kernel; $\Omega_S$ are the exciton energies; and $A_{vc}^S$ are the corresponding amplitudes. We used a plane-wave cutoff of 50 eV for the local-field components of the electron–hole interaction and a scissor shift of Δ = 0.9 eV applied to the single-particle energies[60]. The BSE Brillouin-zone sampling was Γ-centered 45 × 45 × 1 mesh and the exciton basis contained six valence and six conduction bands around the gap, while 400 bands were used to build the screening/local-field matrices. A two-dimensional Coulomb truncation was adopted to account for the slab geometry.

To include a perpendicular external electric field $V^z$, we considered two approaches: an explicit treatment at the Kohn–Sham level[53] and a perturbative scheme at the exciton level. For computational efficiency, we adopted the latter. To first order, the exciton Hamiltonian becomes:

$$H_{SS'} = \Omega_S \delta_{SS'} + V^z D_{SS'}^z \tag{2}$$

where $D_{SS'}^z$ is the out-of-plane excitonic dipole matrix element:

$$D_{SS'}^z = \sum_{vcc'} A_{vc}^{S*} A_{vc'}^{S'} \langle c|\widehat{d_z}|c'\rangle - \sum_{cvv'} A_{vc}^{S*} A_{v'c}^{S'} \langle v'|\widehat{d_z}|v\rangle \tag{3}$$

Here $\langle c|\widehat{d_z}|c'\rangle$ and $\langle v'|\widehat{d_z}|v\rangle$ are the single-particle out-of-plane dipole matrix elements within the conduction and valence subspaces, respectively. The dipole matrix elements were evaluated including PAW corrections.

The field dependence of the exciton energies $\Omega_S(V^z)$ and amplitudes $A_{vc}^S(V^z)$ was acquired by diagonalizing Eq. (2).

## DATA AVAILABILITY

The data that support the findings of this study are available on Zenodo at doi: XXXXXX.

## FIGURES

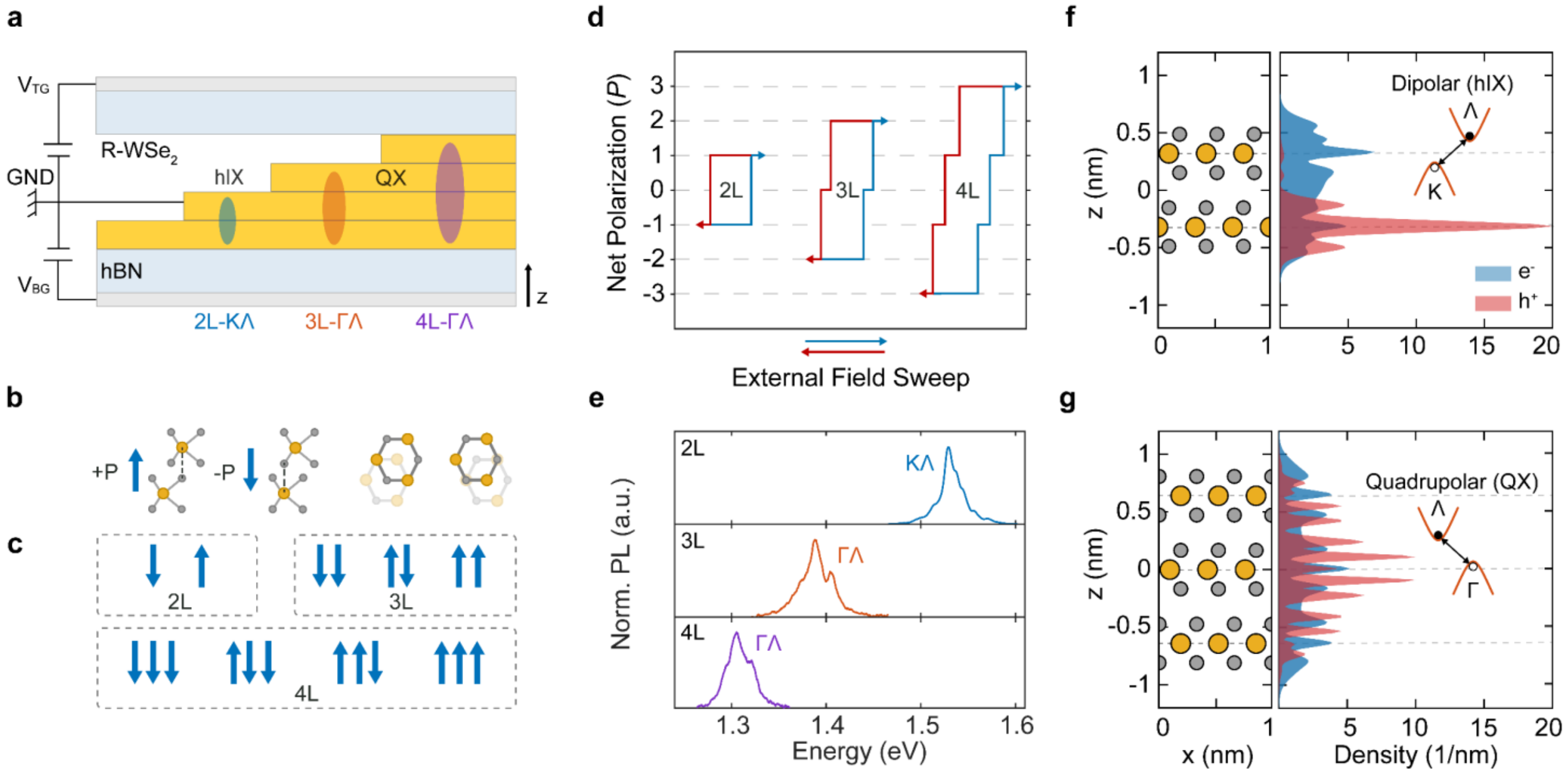

**Figure 1. Excitons and sliding ferroelectricity in few-layer R-$WSe_2$.** **a,** Device schematic of our artificially R-stacked $WSe_2$ (bi-, tri- and tetralayer), encapsulated with hBN and dual-gated for electrostatic control (device micrograph in Supplementary Note 1). We illustrate the different interlayer exciton species, namely hybrid dipolar excitons (hIX, blue) in bilayer and quadrupolar excitons (QX, red and purple) in tri- and tetralayer. Their band transitions are labelled below, specifically KΛ in bilayer and ΓΛ in tri- and tetralayer (DFT, Supplementary Note 3). **b,** Spontaneous electrical polarization (±*P*, blue arrows) of the two metastable atomic configurations arising from R-stacking of two $WSe_2$ monolayers. The two configurations are metal (yellow) above chalcogen (grey) (left side-/top view) and chalcogen above metal (right side-/top view). **c,** Different combinations of polarization with layer scaling in few-layer R-$WSe_2$. Arrows represent electrical polarization at each interface. **d,** Hysteresis loops illustrating the evolution of net polarization through forward (blue) and backward (red) electric field sweeps with layer scaling. Each step is $|2P|$ and relates to the sliding at one interface. **e,** Normalized photoluminescence (PL) spectra of the interlayer excitons observed in bi-, tri- and tetralayer with their respective band transitions. **f-g,** (Left) Atomic configuration for bilayer (f) and trilayer (g) R-$WSe_2$ with net polarizations $+P$ and $0P$, respectively. (Right) Real-space charge distribution of electron (blue) and hole (red) for the atomic configurations shown in the left panel. Wavefunction distributions lead to hybrid dipolar (2L) or quadrupolar (3-4L) excitons. Insets show the lowest-energy band transitions from DFT.

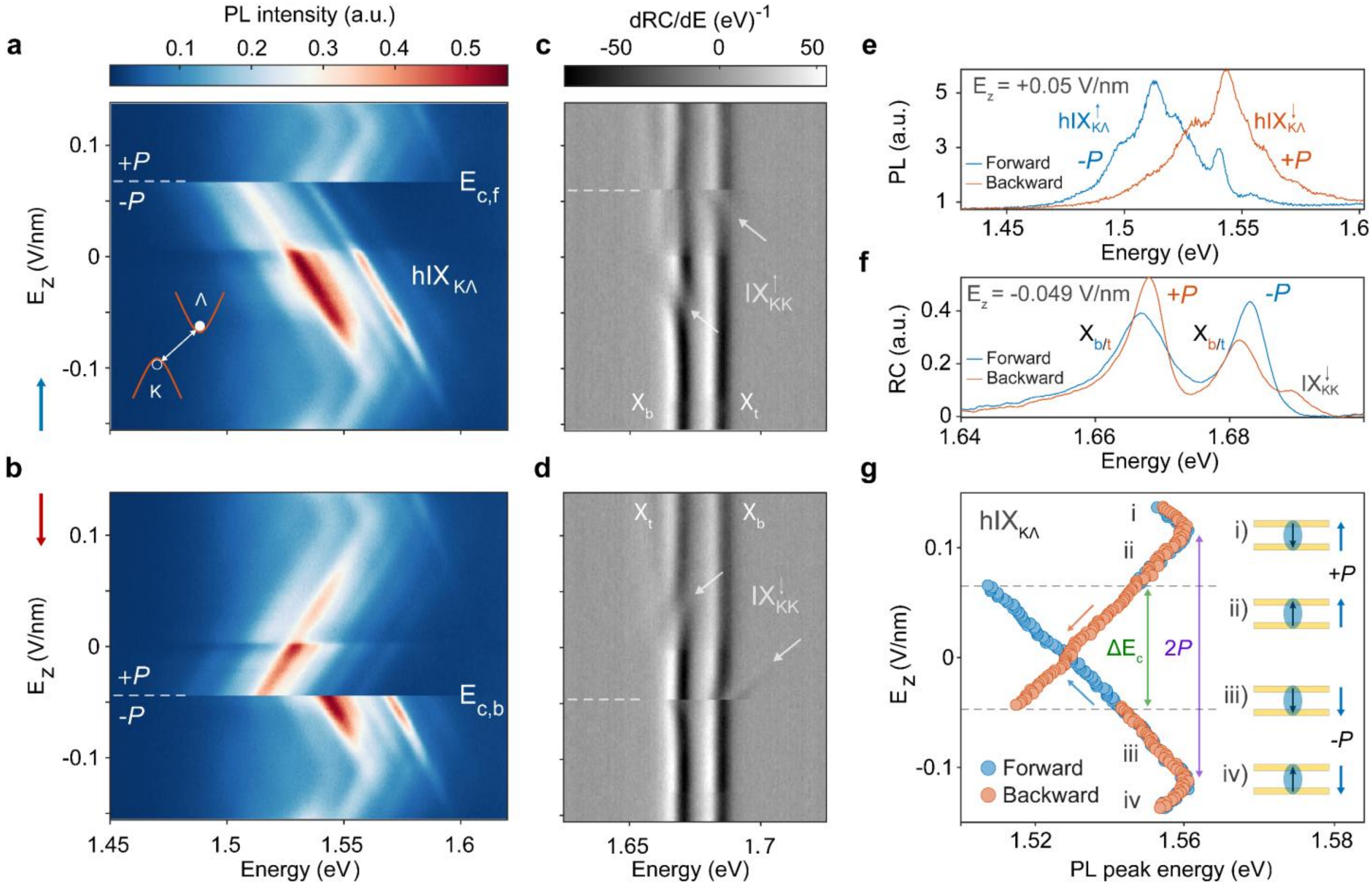

**Figure 2. Spectroscopic signatures of sliding ferroelectricity in bilayer R-WSe₂. a-b**, Field-dependent photoluminescence (PL) spectra in the forward (a, blue arrow) and backward (b, red arrow) sweep direction, obtained by optically exciting our bilayer R-$WSe_2$ with a 640nm CW-laser (Methods). The PL spectra are dominated by bright emission from hybrid dipolar excitons $hIX$ with a KΛ-transition (inset). This transition exhibits a linear Stark shift that either red- or blueshift depending on the field-range and sweep direction. We use the abrupt changes in energy shift and PL intensity as probes for interfacial sliding that switches the polarization between ±*P* at the coercive fields $E_{c,f}$ (forward) and $E_{c,b}$ (backward), marked with white dashed lines. **c-d**, First-derivative reflectance contrast (RC) spectra from forward (c) and backward (d) sweep directions, dominated by non-degenerate intralayer excitons emitting from the bottom ($X_b$) and top ($X_t$) layers. Interfacial sliding is observed at the same coercive fields as in PL, seen from the abrupt changes in intensity. The abruptness is more noticeable due to a pair of anti-crossings between intralayer excitons $X_{b/t}$ and momentum-direct interlayer excitons $IX_{\mathrm{KK}}$ highlighted with white arrows. **e**, Exemplary PL spectra of $hIX_{\mathrm{K\Lambda}}$ at $E_z = 0.05\ \mathrm{V/nm}$ for both sweep directions, depicting a sharp energy shift between $\pm P$ polarization. The hybrid excitons are labelled with superscript arrows (↑↓) to indicate the dipole orientation with respect to the *z*-direction. **f**, Exemplary reflectance contrast (RC) spectra of $X_{b/t}$ at $E_z = -0.049\ \mathrm{V/nm}$, showing intensity re-distribution between intralayer peaks for $\pm P$ polarization. **g**, Stark shift of the hybrid dipolar $hIX_{\mathrm{K\Lambda}}^{\uparrow\downarrow}$, plotted for forward (blue) and backward (red) sweep directions. We extract the energy shift from the maximum PL intensity. We observe a clear hysteresis within the coercive fields (dashed lines), approx. centered around zero field. This leads to a memory window of $\Delta E_c = 0.112\ \mathrm{mV/nm}$ (green double-headed arrow). The polarization switching leads to a field-offset of $0.231\ \mathrm{mV/nm}$ corresponding to $2P$ (purple double-headed arrow). The hysteresis is characterized by four distinct combinations of dipole orientation and polarization, denoted with labels (i-iv) in the extracted Stark shift and schematically illustrated in insets. The exciton wavefunction (blue cloud), the dipole moment (black arrow) and polarization (blue arrow) are sketched.

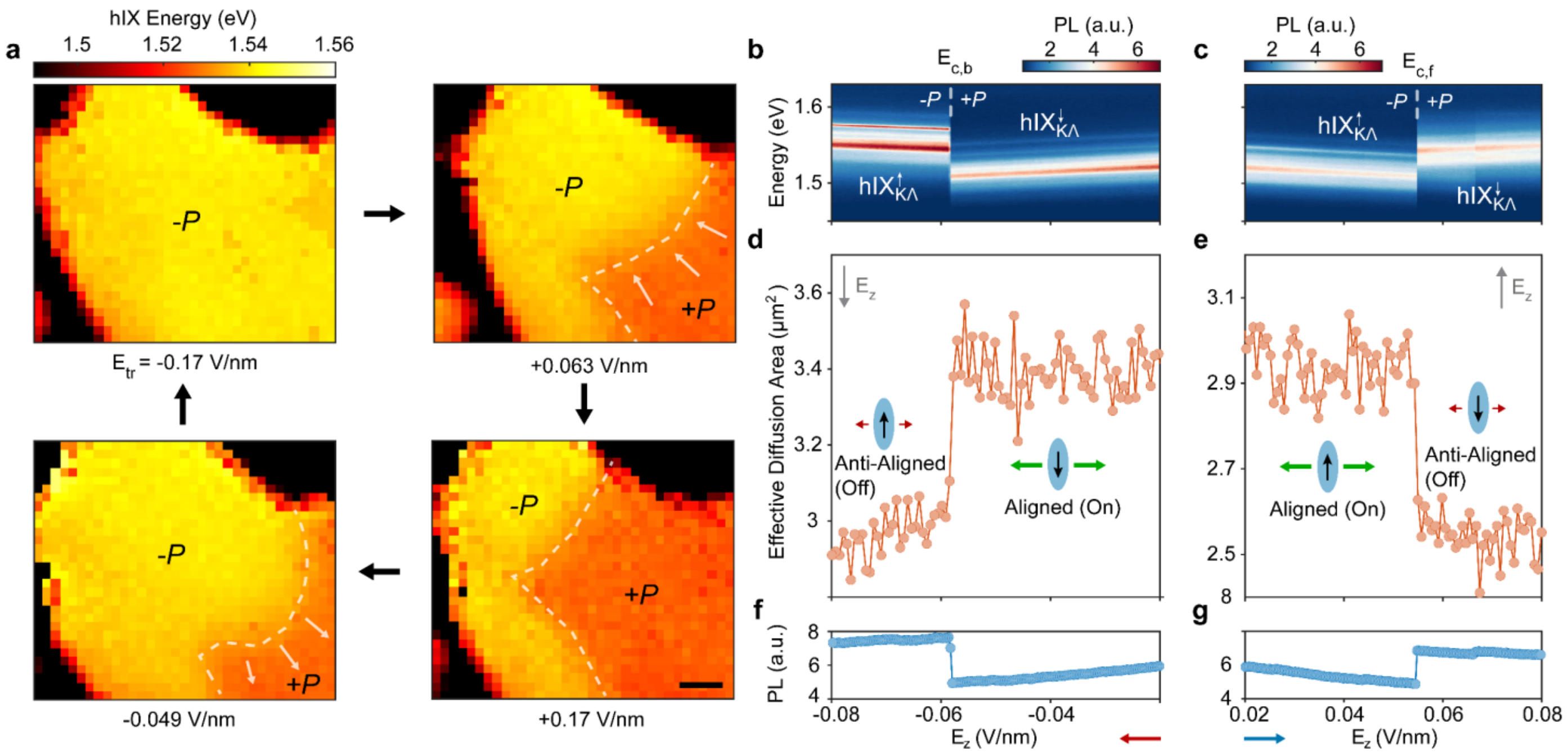

**Figure 3. Sliding-ferroelectric control of polarization domains and exciton diffusion in bilayer R-WSe₂. a,** PL spatial mapping of bilayer R-$WSe_2$ using the peak energy of hybrid dipolar excitons hIX for probing domain wall propagation. We apply various training fields $E_{tr}$ before each map and measure PL at $E_{meas.} = -0.04\ \mathrm{V/nm}$. Domain walls are drawn with dashed lines, and their propagation path is marked with white arrows. The four maps represent a reversible hysteresis loop mapped in the direction of the black arrows. We fit the PL spectrum of each pixel with a Gaussian to extract the energy. Colour scale comprises the observed bilayer energies. Nearby regions are monolayer and trilayer regions (Supplementary Note 1). The maps are measured at room temperature. Domain wall propagation of the entire bilayer region is shown in Supplementary Note 12. Scale bar is 1 µm. **b-c,** PL spectra acquired across the polarization switching for backward (b) and forward (e) sweep directions. We highlight the coercive fields in backward ($E_{c,b}$) and forward ($E_{c,f}$) sweeps with dashed lines. The dipole orientation flips at the coercive fields, given by the red- and blueshift. We thus label the hybrid exciton $hIX^{\uparrow\downarrow}_{\mathrm{K\Lambda}}$ with superscript arrows (↑↓) to indicate the dipole orientation. **d-e,** Effective diffusion area of the $hIX_{\mathrm{K\Lambda}}$ along the backward (c) and forward (f) field sweeps, simultaneously measured with the PL spectra above. The diffusion area is extracted from spatial PL images (Supplementary Note 13), with the threshold of 1/e of the maximum intensity. Insets illustrate the switchable exciton diffusion, ferroelectrically controlled by aligning/anti-aligning the exciton dipole (black arrow) to the electric field (grey arrow), analogous to an on/off switch. **f-g,** Integrated PL intensity of $hIX_{\mathrm{K\Lambda}}$ showing an abrupt intensity increase at the coercive fields, indicating a drastic increase in PL efficiency between aligned/anti-aligned exciton dipoles with respect to the field. This picture is supported by exciton lifetime measurements in Supplementary Note 14.

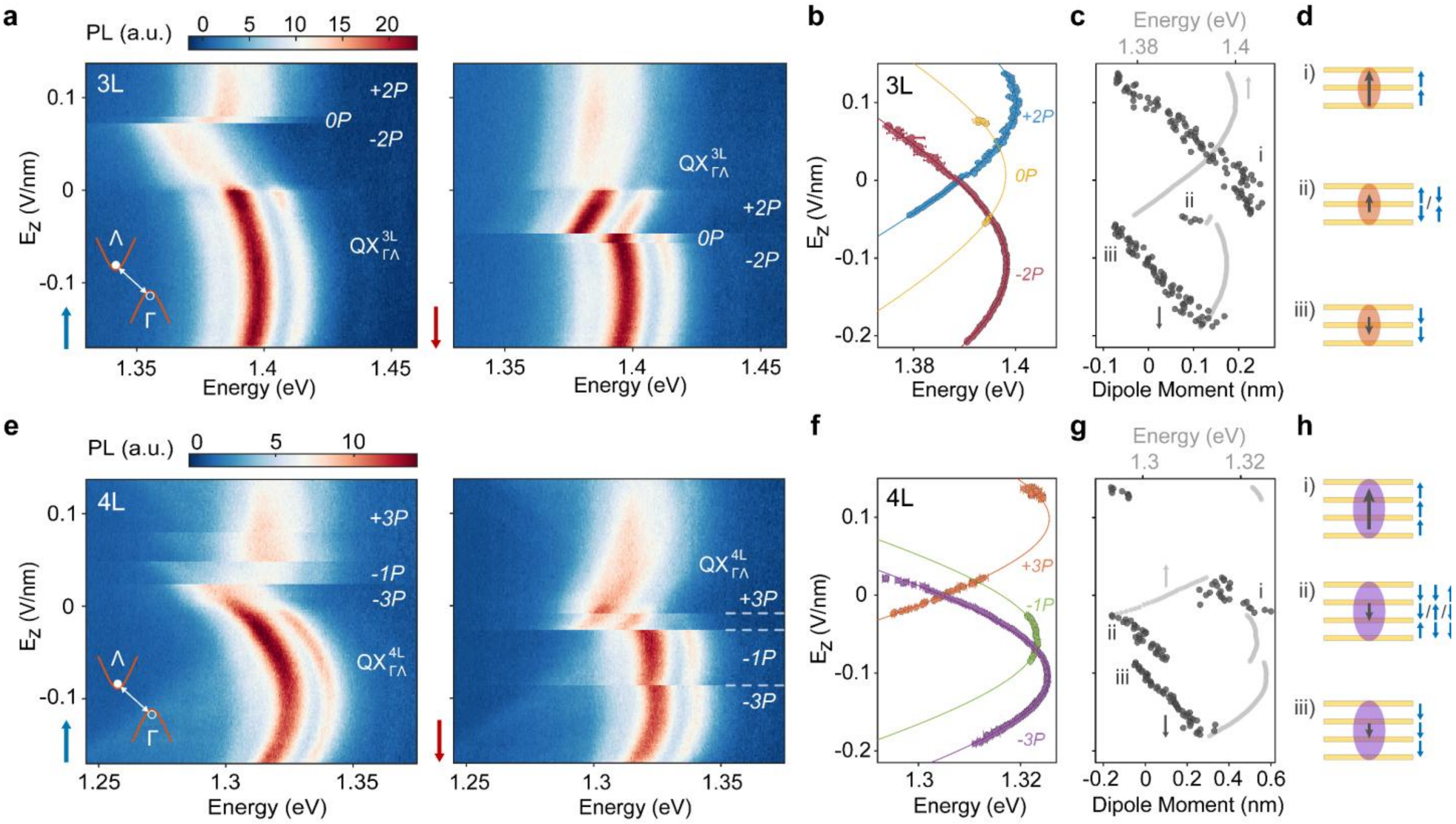

**Figure 4. Quadrupolar excitons and multi-ferroelectricity in few-layer R-WSe₂. a,e,** Field-dependent PL spectra from trilayer (a) and tetralayer (e) in the forward (left) and backward (right) sweep direction. Quadrupolar excitons dominate the field-dependent spectra with low-energy emission for both trilayer ($QX_{\Gamma\Lambda}^{3L}$) and tetralayer ($QX_{\Gamma\Lambda}^{4L}$), originating from the momentum-indirect ΓΛ-transition (insets). Sliding ferroelectricity manifests as a field offset in the QX Stark shift and introduces hysteresis between the forward/backward field sweeps. We indicate the coercive fields (white dashed lines) and label the net polarization based on the field offset in the Stark shift. **b,f,** Extracted Stark shift of quadrupolar excitons from field-dependent PL spectra for trilayer (b) and tetralayer (f), bunched into separate polarization states. We resolve the following polarizations for trilayer $-2P$, $0P$, $2P$ (red, yellow, blue), and tetralayer $-3P$, $-1P$, $+3P$ (purple, green, orange). The plotted energies and associated standard error are extracted from multi-Gaussian fitting (Supplementary Note 16). For simplicity, we only plot the energies of the brightest peak. We fit the energy shift within each polarization regime with the model $\Delta E(E_z) = -\sqrt{t_0^2 + (eE_z \cdot d_0)^2}$ and plot as continuous lines. **c,g,** Dipole moment (dark grey) for trilayer (c) and tetralayer (g) in the backward sweep direction, showing both gradual and abrupt tuning. We apply a noise filter to smooth the energy shift before calculating the dipole moment ($\Delta E = -E_z d$). The energy shift (light grey) for the backward sweep is plotted on the top x-axis. **d,h,** Schematic illustrations of the quadrupolar exciton dipole at specific fields (i-iii) for trilayer (d) and tetralayer (h), showing both gradual and abrupt tuning. Exciton dipoles are represented with black arrows. Polarization at each interface is labelled with blue arrows on the side.